# The Relevance of the Preparation Concept for the Interpretation of Quantum Formalism

**Authors:** M. Ferrero*[1], V. Gómez-Pin[2], D. Salgado[3], J.L. Sánchez-Gómez[4]

**Affiliations:**

1.- Dpto. Física, Universidad de Oviedo. 33007 Oviedo, Spain.
2.- Dpto. de Filosofía. Universidad Autónoma de Barcelona. 08193 Bellaterra. Spain.
3.- D. G. Metodología y Tecnologías de la Información y las Comunicaciones. Instituto Nacional de Estadística, 28071 Madrid. Spain.
4. - Dpto. Física Teórica, Universidad Autónoma de Madrid, 28049 Cantoblanco. Spain.
*Correspondence to: maferrero@uniovi.es

**Abstract**

The preparation procedure, an undefined notion in quantum theory, has not had the relevance that it deserves in the interpretation of quantum mechanical formalism. Here we utilize the concepts of identical and similar preparation procedures to show the conceptual differences between the statistical and the conventional interpretation of quantum formalism. Although the statistical understanding, and its final logical consequence, hidden variables theories (this connexion being explained in the text), have a great intuitive appeal due to its fewer ontological difficulties, both recent experimental results and theoretical developments seem to support an epistemic alternative closer to the conventional one. Nevertheless, we must not rule out the possibility that new theorems or new explanatory principles may modify the reigning supremacy of this interpretation.

## 1.- Introduction.

Although the conventional interpretation of the quantum mechanical formalism (CIQM), very close to the Copenhagen interpretation, is still dominant (is used in almost all textbooks), there are physical situations that ask for an interpretation of the quantum mechanical formalism (QMF) in terms of the Statistical Interpretation of Quantum Mechanics (SIQM) or even for hidden variables theories (HVT). The SIQM and HVT are different research programs, but they are closely related, as we will see further below. To introduce properly the distinctions between the three previous acronyms we will ask an old and apparently simple question. Is the wave function associated to an individual system or to a statistical ensemble? A first tentative answer could be formulated as follows. QMF introduces probabilities, and probabilities are related to statistical ensembles. Otherwise, the same concept of probability would lose its meaning. In QM, if it is said that you would get some concrete result with 0,63 probability it means that if the experiment is carried out 100 times, that concrete result would appear approximately 63 times. Hence, the obvious answer will be that the wave function should be related with statistical ensembles. However, despite its intuitiveness this idea does not capture completely what we want to express. The real problem to analyze could be best introduced with the following example. Imagine that we *prepare identically* two radioactive atoms with half-life of 30 minutes.

By *preparation* we understand all those theoretical methods and physical procedures necessary to fully characterize the state of the system. It is, thus, a macroscopic reproducible act that lefts the system, without destroying it, in a concrete state after the procedure. In quantum theory, this preparation is represented by a vector state (or a density matrix). *Identically* in this context will mean identity of procedures, that is that the experimenter has repeated all previous operations, so that there is no difference whatsoever between them. Note, however, that *identical preparations* are not necessarily followed by *identical results*, so there is an important practical problem in guarantying that the state is the same after the preparation process.

Identity, as we have used it here, is referred to the preparation procedures. However, in physics, the term identity is usually applied to entities, not to procedures. Two or more entities (or individuals) are identical if they have all their intrinsic *properties* in common (and reciprocally). *Identical preparations* do not imply necessarily *identity of entities* (see comment below). One interesting philosophical problem related to the previous distinction, which we will quote on passing, would be if individuation is lost in those circumstances or if we still can speak about "individuals". This is an old philosophical problem with many facets. When in 1663 Leibniz attempted the synthesis of the arguments introduced by his predecessors in relation to the principle of individuation (Quillet, 1979), he took a position close to the nominalists. The reason was that he considered that only individuals existed and, moreover, that the division between two individuals is not a particular feature added to the constituents of the identity, but the expression of the differences between two complete set of those constituents. In other words, entities that share the whole set of features are indiscernibles, then identical. In physics all the particles (or individuals) of the same type, are identical. Electrons, neutrons, protons, etc., produced at different times and places of the universe have the same intrinsic properties (there are not errors in their manufacturing) and, therefore, they are truly identical. Swapping two of them in an ensemble changes nothing, except a possible sign. Being identical they still can be either distinguishable (in classical mechanics) or indistinguishable (in quantum mechanics) [1]

Let us go back to the radioactive case we were considering. Here the synchronic "indiscernibility of the identicals" would refer *only to the preparation* procedures and would resolve diachronically with time. A subsequent controlled experiment would reveal a differential property in the entities: the first would decay, for example, after 16 minutes, whilst the other would decay after 44 minutes, let us say. As we have already said, the identity of preparations do not lead either to the identity of the individuals, nor to the identity of the results.

The CIQM conceives these two atoms as *identically prepared* and adds that *there is no explanation* about why the outcomes are different, that is why one of them decays after 16 minutes and the other after 44. Even more: it defends that such explanation does not exist. Consequently, from this point of view Leibniz's converse principle of the indescernibility of identicals would in some way fail in the quantum domain. We would be in a situation in which, in the absence of a distinctive trait in the preparation, we would have a plurality of identical procedures referring to entities that nevertheless behave differently.

On the other hand, the SIQM would maintain that if the decay takes place at different times it had to be because the *preparation* procedure does not give a homogeneous ensemble, an ensemble of objects ontically identical, but *similar* and that the two atoms must have, therefore, some differential intrinsic properties. This is close to the idea we want to explore, namely that statistical ensembles can be *identically or similarly prepared* and that the concept of identical preparation is related to the CIQM, while the concept of similar quantum-state preparation is connected with the SIQM and, perhaps, with HVT. This has other implications that we will disclose in what follows.

Note that we are not dealing here with the related philosophical problem of identity and individuality already mentioned (French, 2011). We are dealing only with the problem of the meaning of the concept of preparation (identical or similar) and its consequences as far as the understanding of quantum formalism is concerned. Unlike the SIQM, where the referent of the state preparation is a statistical ensemble of microscopic objects, in the CIQM it is understood that the state is a representation of the result of a preparation procedure, rather than an ontic description of microscopic objects.

---

[1] We have omitted in this brief digression the spatio-temporal properties. In classical mechanics all identical particles (atoms, molecules, electrons, etc.) are considered *distinguishable* by the different spatio-temporal properties (positions and velocities) that they may have at a given time. However, in quantum statistical mechanics it is postulated and well established experimentally that a particular type of identical particles may be indistinguishable from one another, as the different statistics shows. Classical statistics, based on the distinguishability of identicals by their spatio-temporal properties, is called Maxwell-Boltzmann statistics, while in quantum mechanics the indistinguishability of identicals may result in the Bose-Einstein or the Fermi-Dirac statistics.

This article is addressed primarily to students and recent graduates in physics and philosophy, but we pretend it should be also of interests to a broader audience, including philosophers, physics teachers, historians of physics and scientists. It is organized as follows. First, we introduce briefly and schematically the difference between the two quoted interpretations from the point of view of the preparation process. Then, in part three, we discuss if the SIQM leads necessarily to HVT o not, and their relation with local realism. Finally, some conclusions are gathered.

## 2. – Epistemic versus ontic preparation.

For the purposes of this paper, the CIQM will be characterized by two different traits. The first aspect is that the quantum state $|\psi\rangle$ is a mathematical expression that *represents* the result of the preparation procedure, either of an individual system, or of an ensemble of systems composed by two, three… N, particles (if we recursively repeat the procedures). Loosely speaking, we could say that the state of the system $|\psi\rangle$ encodes relevant *information* about those degrees of freedom necessary to completely characterize the system in a *defined experimental context*. The second aspect, related with the first one, is the introduction of an arbitrary *division line* between what is considered the "observer" and which is considered the observed system. This is done by applying the formalism only to the system, but neither to the sources, nor to the apparatus that carry out the measurement (see below for a qualification). The preparing and measuring devices are usually described in terms of the operations that the technicians, that have previously calibrated them, "know how" to do to make the experiment. Naturally, this procedure splits the world in the two mentioned parts: the (state of the) system, to which quantum theory applies, and the rest of the world, in particular, the measurement apparatus, to which the theory does not apply.

Although the previous description coincides more with Bohr's particular approach (Scheibe, 1973), historically there have been two different sensitivities to understand where to place this division line inside the supporters of this approach.

a) The first sensitivity[2] maintains that it is a matter of complete indifference where the separation line is situated. We have to put the line somewhere, but this is not really cause of concern as far as contradictions do not arise[3]. Quantum theory has the same meaning wherever we put the line on. Let us explain this point with one example. Imagine that a person in a laboratory is measuring the spin projection over some direction of a spin ½ particle with a Stern-Gerlach device. For her, the particle is on one side of the division line and the Stern-Gerlach and herself, on the other. Suppose now that another person is watching what is going on in that laboratory through a video camera. This second person can legitimately consider that the complete laboratory, including the first person and her devices, is her system. Now the line has been shifted. The whole laboratory, including the first person, is on one side and her, the new observer, on the other. They are describing different systems. We would have a serious problem with quantum formalism if the first observer predicts with certainty spin up and the second one predicts with certainty, and for the same measurement direction, spin down. However, there would be any physical problem at all if the two observers always coincide in their predictions. This issue has been analyzed a few years ago and it has been found that, in fact, there is no contradiction at all: the shifting of the line is not indeed a problem in this sense. Unfortunately, lack of space does not allow us to deepening into this discussion. The interested reader should consult Peierls (Peierls, 1991) and Brun (Brun et al, 2002, and references therein).

---

[2] Related to the names of N. Bohr, M. Born, W. Heisenberg, W. Pauli and P. Dirac.
[3] This problem seems to have some interesting similarities with the proposal of the *extended mind* (Clarke and Chalmers, 1998. See also Menary, 2010). It is beyond the scope of this essay to develop this conjecture.

A different matter inside this variant is to establish who the *subject* that marks the aforementioned separation line is. We have just shown with the previous example that different observers can assign different state vectors (or wave functions) to the same system depending on the information they have. Consequently, the state vector (alternatively, the density matrix $\hat{\rho}$) seems to encode the information that an *individual observer* has about the result of an experimental process classically describable, understanding by this the *contextual information* about some degrees of freedom of the system. This allows two observers that have different information about a concrete experimental situation, to use different wave functions. This individual subjective understanding is strongly supported by many experiments carried out in the last twenty years, the experimental verification (Boschi et al., 1998) of the quantum teleportation protocol (Bennet et al, 1993) being one of them. What is teleported is the information acquired in the *preparation* that, conveniently applied to a distant system, reproduces the initial quantum state on that quantum system. As we have explained above, one of the implications of this point of view is that the division line is inevitably traced by some individual observer, and therefore, the observer necessarily must be taken into account: it cannot be eventually removed. This individual observer might be in a first instance a mechanical device, but in the end, there must be an *interchangeable conscious observer* able to register the result and to incorporate it to physics, to the cultural heritage. Undoubtedly, experiments have outcomes, but the outcomes lie outside quantum formalism. This is a crucial consequence of the quantum postulates whose transcendence, unfortunately, is often overlooked. To discuss it with detail lies outside the purposes of this article and the interested reader should consult Mittelstaed, 1998. What the QMF allows in general is to predict the outcomes with certain probability, and not to determine them through the interaction between the microscopic system and the measuring device. There is not such a physical description in the QMF. Quantum results are comprehended as real phenomena that cannot be *reduced* to physical processes, as is the case in the Lorentz contraction. They are "experienced objective realities", whatever this would mean, and they should be considered as *primitive* non-explainable facts to be subsequently organized by the theory. This is the ultimate reason why the line is necessary.

This individual observer plays thus the role of a transcendental subject, limited to the cognitive operation, and not the role of an aesthetic or ethical subject. In the sense we are speaking here, ontology is hardly dissociable from the theory of knowledge: it is the intervention of the transcendental subject what shapes the objectivity of things. This is the reason why the individual observer must "communicate to others what she has perceived and what she has learnt" (Bohr, 1958). *Events known by nobody, are not part of physical science*. This scrambling of realities and information (Jaynes, 2003) is perhaps the essential characteristic of this first variant of the conventional interpretation (the CIQM).

b) The second sense or variant of this interpretation applies the formalism also to the devices but without any *apparent* separating line. Von Neumann first intended this bold step, which would restore a homogeneous account of the world based on the quantum principles, in 1932 (von Neumann, 1955). However if we make this step, we could find ourselves trapped in the same situation we try to avoid by introducing the arbitrary line as a primitive concept. Let us explain briefly this point. If we apply the formalism to both, the system and the apparatus, the linearity of the theory has the consequence that the system and the measuring apparatus become entangled. This means that the joint state corresponds to a superposition of at least two empirically *incompatible* macroscopic possibilities. Both terms are present by virtue of the unitary evolution and no result will be produced. Technically, pure states evolve unitarily to pure states but to get a result we need one of the terms in a mixture. Entanglement is then only the precedent of what we call a measurement. In the entangled state, we still do not have a result. To get a result we need to determine which one of the two possibilities, let us say, is actualized by measuring. If we had an irreducible separating line, we could appeal to the classical character of the apparatus and decree that the pointer has always a definite position, as the first variant maintains. The position of the classical pointer selects only one term in the superposition, and the problem disappears (by definition). As we have said, this is how Bohr solved it. However, if there were not a separating line, as required by those looking for a homogeneous account of the world, the problem would actually remain. A possible way out would be to extend the system with another observer that watches what is going on in the second laboratory, and them with another one, and so on (von Neumann's chain), until you reach a point in which, to have the desired outcome, "something" must reduce the mentioned superposition, leaving only one term. What can be then this "something" that reduces the superposition to a single term producing a result? *It cannot be anything physical*, because, in that case, we could recursively apply again the argument. One possibility left is that, if we wish to have a coherent scheme, entities able to produce projections, should exist. The best candidates to do that action are an external God that intervenes on the universe (not an Aristotelian, but a Christian one) if the system is the whole universe, or, alternatively, the conscious mind of the observer (in all the other cases). The first alternative could be considered as a kind of "quantum ontological argument" for the existence of God and lies outside the objectives of this essay. The second introduces the mind-matter problem and, in this sense, also lies outside the objective of this essay. This alternative, with strong similarities with the previous one, corresponds, roughly speaking, to what has been called the von Neumann (v. Neumann, 1955), London-Bauer (London and Bauer, 1939) and Wigner (Wigner, 1967) interpretation.

Note that this approach is not really different of the first mode above: there is only a shifting of the line, not a real elimination of it. The line has now an objective place: it will be always situated in the mind of at least one observer that perceives the result, a conscious observer that must, afterwards, *communicate to others* what she has learned. This communication process is not an irrelevant step, but absolutely necessary to incorporate that result to physical science.

This interpretation, considered here as a second CIQM variant, and the Platonic (or the Cartesian) dualism introduced, is a fascinating subject by itself, more if we consider the proposals put forward in the last twenty years by researchers at the interface between consciousness and quantum theory (Stapp, 2001 and 2007). Textbooks combine these two understandings of QMF and this somewhat confusing mixing is usually called the conventional interpretation of QMF[4]. We think that it is possible to unify coherently these two interpretations (leaving aside any mystical connotation) if the emphasis is put in the two following aspects already mentioned. First, the state of the system should be related to the acquisition of contextual information that takes place in the preparation procedure. And second, the process of obtaining a result should be understood not as a *physical process*, but as an epistemological one, that is, as the acquisition of information by science, physics in this case. Unfortunately, lack of space does not allow us to develop more the coincidences and differences of these two variants and we leave it for a subsequent paper.

The relevant conclusion for our purposes, to be to drawn from the previous comments, is that the association of the state vector (the density matrix) with identical preparation procedures should be related the conventional interpretation of the QMF and, as we have tried to show, it is inextricably linked to the so called measurement problem (Albert, 1994) (and perhaps to the problem of the conscious mind. See Rosemblum et al., 2006; Squires, 1990).

On the other hand, in the SIQM it is considered that $|\psi\rangle$ *corresponds* to the real state of an ensemble of microscopic objects. However, due to "practical inevitable circumstances"[5], the preparation procedure of this state is always imperfect and thus the state vector must be related exclusively to (an ensemble of) *similar prepared* systems. In this interpretation, the state has an objective meaning and corresponds to the ensemble. For the SIQM, QMF remains silent as far as individual systems are concerned. It speaks exclusively about ontical statistical ensembles (Ballantine, 1970 and 1998), the statistical character of quantum mechanics being related to incomplete information of the individuals. In the CIQM, the statistical character of the theory is however due to the essentially probabilistic nature of physical world itself.

---

[4] Some scholars call it "the shut up and calculate interpretation" (SUCI).
[5] To prepare identically an ensemble of systems is considered here literally pure metaphysics.

Note that, regardless of what we have argued, to contend that the two atoms are "really identical" is somehow going beyond the CIQM. As we have already said, the identity of the preparation procedures does not imply in any way the identity of the alleged entities. The reason is that the CIQM sees the theory as formulated in terms of knowledge, a philosophical combination of pragmatic (Rorty 1999) and structuralist (Ludwidg, 2008) elements. As it is well known, and Niels Bohr emphasized many times, the CIQM does not speak about ontic essences in any way, but about the result of our potential future interventions carried out in concrete experimental macroscopic contexts and about their mutual relationships.

Although the conventional interpretation of QMF has been proved consistent and productive for the last 87 years, there are, however, situations in which not considering the quantum state as corresponding to a real quantum system puts strong pressures on our understanding of what is going on, pressures that will become evident with the following example. Let us consider an ethanol molecule[6]. If we ignore the spin, that is not relevant for our considerations, quantum theory tells us that the total angular momentum in the ground state of this molecule is equal to zero. What does it meant to say that the angular momentum of an ethanol molecule equals zero? Although this is a difficult problem that requires a complex technical treatment in which we cannot enter now, let us say that according to the CIQM, this means that the state has *spherical symmetry*. Now, if the state represents the knowledge we have about the preparation procedure of an individual system, what does it mean to say that this knowledge has spherical symmetry? Would it imply that one individual $CH_3 - CH_2OH$ molecule has spherical symmetry?

A much more easy way to comprehend this symmetry would be to think that the wave function corresponds to a statistical ensemble of molecules *similarly prepared*, as the SIQM maintains. In *one* of these molecules, the axis that passes through the two carbon atoms, with black colour in the figure below, would point, for example, horizontally in relation with some arbitrary coordinate system; in other, would point at 45º; in another one at 31º; and so on, with an isotropic distribution. If that were the case, then it would be meaningful to say that the whole ensemble, the state referent for the SIQM, has spherical symmetry. However, as far as one individual molecule is concerned it is impossible to figure out that symmetry. The reason is made evident by looking at the figure 1.

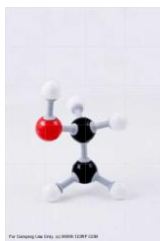

Figure 1. Ethanol molecule.

Two carbon atoms are located in the centre of two separated tetrahedra, inverted and joined at the vertices. The angle formed by the line that joins the two carbon atoms and the one that joins the oxygen with one of the carbon atoms, has been measured to be about 103º. It also has been measured with high precision other angles, the strength of the links between the different atoms, etc. To summarize: we *experimentally* know that the molecule has the structure modelled in figure 1. If we think, as the SIQM does, that QMF is formulated in terms of a "finished" objective reality[7] and that the sate vector directly corresponds to this reality, it would not be possible to conceive the spherical symmetry of one individual molecule in any way. It has no meaning at all. This is one the reasons why, in this interpretation, the referent of the quantum state is the ensemble. The counterpart is that, like in Aristotle, we only know things about ensembles; the individuals are not object of science. In brief, the ontic meaning of the quantum state defended by the SIQM is well founded experimentally. It could be understood as a physical object encoding all the properties of this molecule. However, practical limitations may prevent one from *preparing identically* the molecules such that when the preparation procedures are finished, what we get is an isotropically distributed ensemble. The spherical symmetry would be then easily understood if we associate the wave function with a conceptual Gibbs ensemble, and not with an individual ethanol molecule. In that case, the different individuals of that ensemble will not be *identically*, but *similarly* prepared, as the SIQM advocates. With an ensemble of identically prepared systems, it would not be symmetry. If they were identical, all axes should point in the same direction and the angular momentum will not be zero.

---

[6] The water or carbon dioxide molecules could be other examples. The latter has the advantage that the three nuclei have spin zero and it is only necessary to justify the zero coupling of electron spins.
[7] That is, with a well defined structure, properties, and so on. More concretely: finished in ontological classical terms.

Note that if QMF is formulated on terms of knowledge, as we have said the CIQM does, the alleged entities do not need to be identical. It is a dubious way out, because there is still an important problem remaining here. This interpretation would be yet in hard trouble for the reason that it is very difficult to escape *the force of the experimental results* in which the structure represented in this figure is based.

If the individuals are similar but not identical, a new possibility not contemplated till now opens: it might be possible to introduce additional underlying variables able to explain the hypothetical differences between the individuals, giving therefore a deeper and more detailed picture of the world, perhaps a classical one. The theories aimed to introduce these variables, not contemplated in quantum formalism, are called hidden[8] variables theories (HVT).

Some scholars make an interesting distinction between the SIQM and the HVT perspective that is convenient to remember now. Both approaches have in common the idea that the wave function does not contain complete information about the individual quantum system, as Einstein pointed out many years ago, but about an ensemble of similarly prepared quantum systems. Despite this coincidence, some of the followers of the SIQM maintain that to introduce additional variables to complete QMF, describing so the individual system, is a *step further* and can be consider a matter of opinion. The reason is that even if the quantum mathematical description of the ethanol molecule were *incomplete*, it might have no meaning to introduce additional parameters, angles for example in this case giving the orientation of the molecule, because it could result in being absolutely useless. It may well happen that we will never be able to produce an ensemble of molecules, for whatever reasons, in such a way that all of them have their axis horizontal, let us say. That is, that they were identically prepared from the ontical point of view. It is indeed easy to argue in favour of this practical impossibility using QMF. It could happen that due to the Heisenberg principle, for example, any preparation that we attempt would result in a spherical distribution of the axis connecting the two carbon atoms. Alternatively, perhaps we could prepare identically all systems as far as *one concrete observable* is concerned, but not in relation to all, in particular the non-commuting ones. In this case, HVT would be useless *in practice* and, from this practical point of view, there is a big difference between the SIQM and (to believe in) the existence of HVT. Many scholars support the SIQM because they see it as the less problematic ontological interpretation of QMF, without considering important the problem of whether to introduce HV or not. In fact, many of them are against the introduction of hidden parameters or ontic properties. This strategy may be considered as a good "protective barrier" against the well-known failures of hidden variables theories, materialized in the so-called non-go theorems and their experimental verifications (see further bellow). To make the HV additional step we must be interested also in the question of if the HV could exist *in principle* and not only if it is convenient to introduce them *in practice*. We will discuss this concrete point in the next section.

---

[8] Hidden for QMF.

## 3. – Determinism, SIQM, Hidden Variables and Local Realism.

Although it is not our purpose to discuss now this issue at length, let us comment briefly that the problems previously considered are related also with *determinism*. QM is not a deterministic theory, as we have explained with the experiment with the two radioactive atoms. It does not predict when exactly an atom is going to decay. It only gives the average life. If somebody prefers, for philosophical or any other reasons, a deterministic theory she should specify things better, as to be able to predict in which instant a concrete atom is going to decay and in which direction the $\alpha$-particle, let us say, will be emitted. This would be another important reason to declare QMF incomplete and to introduce additional parameters. Maybe Einstein had this idea in his head, we will never know. What we do know is that this determinism issue was not his main concern with QMF, but *local realism* that we will introduce now. In this essay, *realism* in the broad sense will mean that there is a material reality, and in the restricted sense, that this material reality is composed by *separable* objects that have dynamical non-contextual, ontic or intrinsic properties that *individualizes* them. *Locality* in the broad sense will mean that actions or influences cannot be propagated instantaneously and, in the restricted sense, that even if these hypothetical influences exist, they cannot be utilized to send messages at superluminal velocity (*no signalling condition*).

To see better the implied connection between the CIQM, HVT, Local Realism and the SIQM, let us remember the example introduced by Einstein in the Solvay meeting (Solvay, 1928). More elaborated, this example took him, jointly with Podolski and Rosen, to the famous EPR argument (EPR, 1935).

Suppose that we have a radioactive atom in point C, the centre of the figure 2 below, and let us imagine that the total angular momentum of the atom is zero. The figure depicted represents an empty sphere whose interior is like a TV screen or a photographic film. If the atom emits an $\alpha$-particle, we will see a sparkle afterwards in the screen or, alternatively, the film will record one event. QMF represents the relevant degrees of freedom of the $\alpha$-particle, after the atom has decayed and before it reaches the screen, by a spherical wave. This

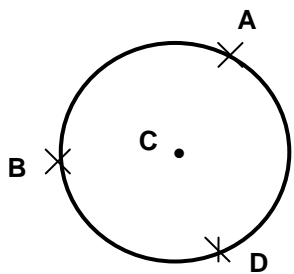

Figure 2. Slight modification of Einstein's gedankenexperiment of 1927.

means that the amplitude of the wave in all radial directions is the same at the same distance of C. However, when the impact takes place and we see a flash on the screen, or a mark on the film, the wave function "jumps", that is, the delocalized wave function, extended by the entire sphere, suddenly "reduces" to the point A in the figure, let us say. Here we have again the two different aforementioned possibilities, (already noted by Einstein at that time):

The first, adapted to our local realist classical intuitions, is to say that the particle has really gone from the centre to the point A, but given that our initial information was incomplete, perhaps due to an impossible perfect preparation, we didn't know to which concrete point of the sphere the $\alpha$-particle was going, and therefore we represent this incomplete information by a function with spherical symmetry. Note that if this realist understanding were applied to an individual case (CIQM) and not only to the ensemble (SIQM), the reduction would imply the instantaneous transmission of "something" physical, as Einstein remarked, because the function that differs from zero at points B or D, for example, suddenly becomes zero on them, when the $\alpha$-particle manifests itself only in point A. If the state vector were a real object corresponding to something more than mere knowledge or information, as it looks like when speaking about one material $\alpha$-particle, and if we apply it to an individual systems, then there must be an instantaneous transmission that violates locality, in the broad sense that an influence has been propagated at a speed greater than the speed of light. Einstein concluded, then, that the wave function corresponds to the ensemble, not to the individual case.

In the CIQM, a pragmatic approach, the wave function is not a physical object. What QMF does is to relate different observed physical phenomena, here the decay of the atom with the flash in the screen. We represent everything we know about the relevant degrees of freedom of the particle by a wave function, which allows us to calculate the probability that afterwards a flash will be produced in the point A of the screen or in any other point (SUCI). These predictions have been, till now, in perfect agreement with the experimental results, and for the purpose of doing science, they add, this is all we need.

On the other hand, the SIQM would allow the following spatio-temporal explanation of the facts. There is a particle; the particle goes, indeed, one way or the other, but *we cannot prepare it* to the point of knowing which way it will go. Only, on the detection, we finally find out which way it has gone. In this explanation is not the case of any transmission at superluminal velocity. It only says that the preparation encoded in the state of the system does not represent the *complete* real state of things, due to the impossibility of making a preparation able to tell us the instant and direction in which the individual $\alpha$-particle will be emitted. What we have is incomplete information due to this same impossibility.

HVT would be a step further from this, and it would respond to the desire of having *a stronger link* with material reality, embodied in a causal, deterministic and spatiotemporal description of processes in the microscopic domain. Causality in this concrete example should be understood as related to the conservation principles. It would mean that if the $\alpha$-particle is not subject to any potential, it will maintain its initial momentum. The spatiotemporal description would tell us the different positions by which the $\alpha$-particle has passed through, so that the place where the impact occurs is determined. This image seems to be a very natural desire.

Alternatively, the CIQM would tell us that position and momentum are conjugated variables in quantum formalism, and they cannot be defined at the same time. We retroactively could, once we know the point on the screen, calculate them, but this knowledge is useless for the purpose of making new predictions (Heisenberg, 1930). As in many other experiments, we can indeed reconstruct the past, a very relevant epistemological issue, but this knowledge cannot help us to make further predictions about the point in which the next $\alpha$-particle will impinge. So, although QMF allow us to *reconstruct the past* depending on the operations we choose to make now, it refers to the *future*, to an open probabilistic future. Deterministic local hidden variables theories (LHVT), and the related local realism, aims at a description that quantum formalism does not allow, they would conclude.

That being said, it is easy to understand why the adherents to local realism would be in favour of the existence of hidden parameters and of the SIQM. We have already commented that if all that can be predicted is given by quantum formalism, the purpose of introducing LHV is not so clear. Let us insist in this idea with the example of the radioactive atom. With a LHVT, we would aim at building up a new theory able to tell us, for example, in which precise moment each atom is going to decay. This theory would be useful if and only if we were able to prepare an atom that would decay after any desired time interval. But we do not know how to do that and perhaps we will never know. All we can do is to prepare the atom, and then it will decay whenever it wants, so to speak. There is only one doubt in this scenario that we want to remember now. Maybe, if we had such a theory, we would be able to prepare it. That is, it could happen that the same new theory would give us also enough information about how to make the preparation procedure. At the present state of our knowledge, we don't know how to do that. And if we cannot control the preparation, we cannot look for a better relation between the preparation and the measurement result. Nevertheless, it is not possible to advance, before building up a theory, if it is going to be fruitful or not. After Popper (Popper, 1934) it is widely admitted that what can be possibly measured is in many cases determined or at least conditioned by the very theory. It is the theory what will tell us what is measurable and what is not. Remember, for example, the motion of Brownian particles. The relevant magnitude to be measured was not the velocity, as it was thought at that time, but the diffusion coefficient. However, this fact was unknown until Einstein published his theory about the Brownian motion in 1905.

We have motivated sufficiently the reasonability of opting for a SIQM in intuitive "difficult" cases. We have also explained the intimate relation between the SIQM, LHVT and local realism. Also the advantage of introducing HVT for understanding what is going on in the microscopic domain. From the ontological point of view, it would be an excellent solution to the most important foundational quantum troubles. However, as it is well known, the problem is that the more attractive hidden variables theories to reach such objective, the local ones, are *incompatible* with the predictions of quantum formalism. The mere fact of its existence implies that some statistical predictions of QM must be false. To renounce to one of the best theories ever constructed by humankind, is not an acceptable option for the physicist's community[9]. The demonstration of this adverse turn of the destiny is known as "the non-go theorems of HVT". As it is not the main purpose of this paper, we will only quote them for completeness, referring to the interested reader to some excellent papers published on this issue (see for example, Mermin, 1993). The first one, and the more famous, is the von Neumann theorem, introduced in his book on 1932 (von Neumann, 1955). Von Neumann's ban was transgressed by D. Bohm (Bohm, 1952), when he was able to produce a HVT that, supposedly, reproduces all the predictions of QMF at the, for many, *undesirable price of contextuality and non locality*. Bohm's theory reactivated the discussion on the possibility of HVT and his work was followed by the works of Gleason (Gleason, 1957); Bell in 1964 and 1966 (Bell, 1987); Kochen and Specker (Kochen-Specker, 1967); Jauch and Piron (Jauch-Piron,1967), etc. All of them introduced non-go theorems, based on hypothesis less restrictive than the von Neumann's ones, that preclude the more attractive HVT (the non-contextual and the local ones) and, paradoxically, weakened the SIQM, the ontic inclined interpretation of the quantum state that historically provided the conceptual basis which supports those HVT. The best solution to the foundational problems of the theory is so theoretically and experimentally blocked[10], and we have to look in a contextual and non-local direction to solve them (Bohm's theory).

## 4. Conclusions.

In this article we have highlighted the relevance of the concept of the preparation procedure to show the basic differences between the statistical and the conventional interpretations of quantum formalism. The results of our discussions are now summarized.

In the conventional interpretation of QMF it makes sense to speak about identical preparation procedures, but that does not imply the preparation of identical entities. In fact, as the theory is formulated in terms of knowledge, we should not really talk about entities at all, but only about relationships between observed phenomena. Consequently, the quantum state $|\psi\rangle$ is seen as containing *information* about the result of the preparation procedure.

There are many instances in this interpretation in which the recourse to the *projection postulate* is inevitable. These projections do not need and do not have any physical description. They are not physical processes but epistemic ones. Due to this fact, some scholars maintain that Quantum Mechanics so understood *is not a physical theory,* a very strong claim, indeed. Their main argument is based on the idea that a physical theory must provide an explanation of the measurement outcomes, based on the physical interaction between the microscopic system and measuring apparatus. That is, it must explain why the device's needle points, for example, to 1 and no to other number. Quantum formalism does not allow obtaining an outcome (Mittelstaed, 1998) and therefore it is not, in their opinion, a physical theory. For them, Bohm's Mechanics or the Collapse Theories (Ghirardi, 2011), close to the SIQM, are physical theories. This strong opinion is not shared by the overwhelming majority of the scientific community, which firmly believes that the CIQM constitutes an excellent physical theory experimentally contrasted up to a level unprecedented in the history of science.

---

[9] To make this risky step we need a really strong motivation like, for example, finding an internal contradiction between their postulates.
[10] Except closing the two remaining loopholes (locality and fair-sampling) in one and the same experiment.

Meanwhile, in the SIQM, it makes no sense to speak of the identical preparation procedures of the entities that make up an ensemble. All that can be done in practice are similar preparations. The entities that make up the ensemble, always kept irreducible differences between them. The quantum state $|\psi\rangle$ is thus a real object corresponding to an ensemble of similarly prepared microscopic objects, inside which the individuals maintain irreducible differences between them. The experimental reasons of these differences are not clear, but the ansatz is that the *preparation* cannot be made empirically complete to eliminate them. The profound physical reason why this is so is an open question. The projection postulate is now understood as the selection of the corresponding sub ensemble. If carried through to its ultimate consequences, the SIQM will be doomed to introduce hidden variables. HV theories maintain that it has conceptual meaning to introduce *in principle* new ontic parameters beyond the quantum formalism to explain the differences between the hypothetical sub ensembles (of the SIQM). Once the allegedly extended theory had been built up, it may happen that this new theory will tell us if the intended preparation can be really done. However, for some of its supporters, if the hypothetical hidden variables cannot be empirically controlled, it has no sense to introduce them, at least from the practical point of view.

The ontic understanding of the formalism (SIQM and its logical consequence LHVT) has a great intuitive appeal because it possesses fewer ontological difficulties. However, many experimental results of the last 20 years carried out with systems that enter into the apparatus one by one, recently recognized with the 2012 Nobel Prize in Physics, jointly with some theoretical developments of quantum information theory seem to be in favour of the epistemic approach[11]. Nevertheless, we cannot rule out the possibility that new theorems or new explanatory principles may modify the current pre-eminence of the conventional interpretation, but it seems improbable that these future advances may incline the balance towards the statistical alternative.

---

[11] Unfortunately, lack of space does not allow us to quote this experimental evidence and theoretical advances.

Ghirardi, G. (2011). "Collapse Theories". Stanford Encyclopedia of Philosophy.

Gleason, A. M. (1957). "Measures on the closed subspaces of a Hilbert space". J. Math. Mech. **6**. 885

Heisenberg, W. (1930). *The Physical Principles of Quantum Theory*. University of Chicago Press.

Jauch, J. M. and Piron, C. (1967). Generalized Localizability. Helv. Phys. Acta **40**. 559.

Jaynes, E. T. (2003). *Probability Theory. The Logic of Science*. C. U. Press.

Kochen, S., Specker, E. (1967), The Problem of Hidden Variables in Quantum Mechanics, J. of Math. and Mech. **17**, 59.

London, F. et Bauer, E. (1939). *La Théorie de l' Observation en Mécanique Quantique*. Hermann. Paris.

Leibniz, W. (1663). *De principio individui*. French Translation "Disputation métaphyique sur le principe d'individuation". See Quilet (1959).

Ludwig, G. and Thurler, G. (2008). *A New Foundation of Physical Theories*. Springer. Berlin.

Ludwig, G. (1985, 1987), *Foundations of Quantum Mechanics I* and *II*. Springer. New York.

Menary, R. Ed. (2010). *The Extended Mind*. The MIT Press. Cambridge. Mass.

Mermin, D. (1993). "Hidden variables and the two theorems of John Bell". Rev. Mod. Phys., **65**, 803.

Mittelstaed, P. (1998). *The Interpretation of Quantum Mechanics and the Measurement Process*. Cambridge U. Press.

Peierls, R. (1991). *More Surprises in Theoretical Physics*. Cambridge U. Press.

Popper, K., R. (1934). *Logik der Forschung*. English translation, *The Logic of Scientific Discovery*, Routledge, New York (1959).

Quillet, J. (1979), "Disputation métaphysique sur le principe d'individuation", Les Études Philosophiques, **1**, 79.

Rorty, R. (1991). *Philosophy and Social Hope*. Penguin. London.

Rosenblum, B. and. Kuttner, F. (2006). *Quantum Enigma. Physics encounters Consciousness*. Oxford University Press.

Scheibe, E. (1973). *The Logical Analysis of Quantum Mechanics* (International Series of Monographs in Natural Philosophy, Volume 56). Pergamon. Oxford.

Solvay (1928). *Électrons et Photons*. Gauthier-Villars. Paris. (In French).

Von Neumann, J. (1955*). Mathematical Foundations of Quantum Mechanics*. Princeton Univ. Press. Princeton, N.J. (German original, 1932).

Squires, E. (1990). *Conscious Mind in the Physical World*. Adam Hilger. Bristol.

Stapp, H. P. (2001). "Quantum Theory and the Role of Mind in Nature". Found. Phys., **31**, 1465.

Stapp, H. P. (2007). *Mindful Universe.* Springer.

Wigner, E. (1967). *Symmetries and Reflections: Scientific Essays*. Indiana University Press. Bloomington.